\documentclass[reprint,amsmath,amssymb,braket,aps,prb]{revtex4-2}
\usepackage{graphicx,color,braket}
\usepackage{dcolumn}
\usepackage{bm}
\usepackage[colorlinks=true,linkcolor=blue]{hyperref}
\usepackage{colortbl}

\usepackage[T1]{fontenc}

\begin{document}

\title{Lamb Shift of Landau Levels in Two-Dimensional Electron Systems in a Multimode Resonator
}

\author{Alexander Shabanov, Georgy Alymov, Dmitry Svintsov}
\affiliation{MIPT, Phystech, Dolgoprudny 141700, Russia}
\email{Shabanov.av@phystech.edu}

\begin{abstract}
The use of resonators to modify the behavior of electromagnetic systems demonstrates its potential for application in a wide range of problems. However, existing theoretical studies often resort to the single-mode approximation, rarely considering a second resonator mode. In this paper, we show that including a large number of resonator modes in the model significantly enhances the softening effect of the cyclotron frequency of a two-dimensional electron system. We address this problem by demonstrating the possibility of reducing the system to a set of coupled harmonic oscillators and finding the eigenfrequencies of the oscillators. This is made possible by applying the self-energy method for modes in one polarization and the method for finding the eigenvalues of matrices that have undergone first-rank updating for modes in the perpendicular polarization.
\end{abstract}

\maketitle

One of the main problems in quantum physics is describing the behavior of electrons in a constant magnetic field, which leads to the formation of Landau levels. This model provides a good understanding of the quantum Hall effect, explaining the quantization of the Hall resistance ~\cite{stone1992quantum}. One way to further influence electron transport in a sample is by distorting the vector potential of the system ~\cite{session2025optical}by placing the system in a resonator. This leads to a change in the spectrum of virtual photons, similar to the Lamb shift, and shifts the eigenenergies of carriers. This problem has attracted considerable attention since the successful observation of a modified Hall effect in the experiment ~\cite{appugliese2022breakdown,enkner2025tunable} due to its ability to control transport phases in materials using optical cavities containing a two-dimensional system ~\cite{lu2024cavity,keren2026cavity,sentef2018cavity,ciuti2021cavity,papic2011tunable,paravicini2019magneto}. This led to the development of a model showing a weakening of the topological security of edge states due to softening of the cyclotron frequency when it hybridizes with the cavity eigenmodes in a model that takes into account a system of several electrons and a single-mode cavity ~\cite{rokaj2023weakened,yang2026quantum,rokaj2024topological}.

However, in solving this problem in this paper, we use a single-mode approximation, which studies the effect on transport of only the resonator mode closest to the Landau level~\cite{bacciconi2025theory,cardoso2026cavity,de2022magnetic,yang2026quantum,rokaj2023weakened,li2018vacuum}. This approximation fails for smaller resonators, which simultaneously achieve a high interaction strength between virtual photons and carriers and new phenomena arising from the collective excitation of modes~\cite{scalari2012ultrastrong,garcia2021manipulating,winter2025fractional,saez2023can}. In later studies(~\cite{tay2025multimode}), models were developed for a multimode resonator for a one-dimensional electron system ~\cite{hagenmuller2010ultrastrong} and a multimode resonator whose mirrors are parallel to a two-dimensional system (~\cite{andolina2026quantum,rokaj2022polaritonic,rokaj2019quantum}). A system consisting of a single-mode resonator and several two-dimensional systems (~\cite{mornhinweg2024mode,de2021light}) was also studied, but this work did not reach a logical conclusion by generalizing its conclusions to the case of an arbitrary number of resonator modes with an arrangement consistent with experiment.

\begin{figure}[ht!]
\includegraphics[width=0.9\linewidth]{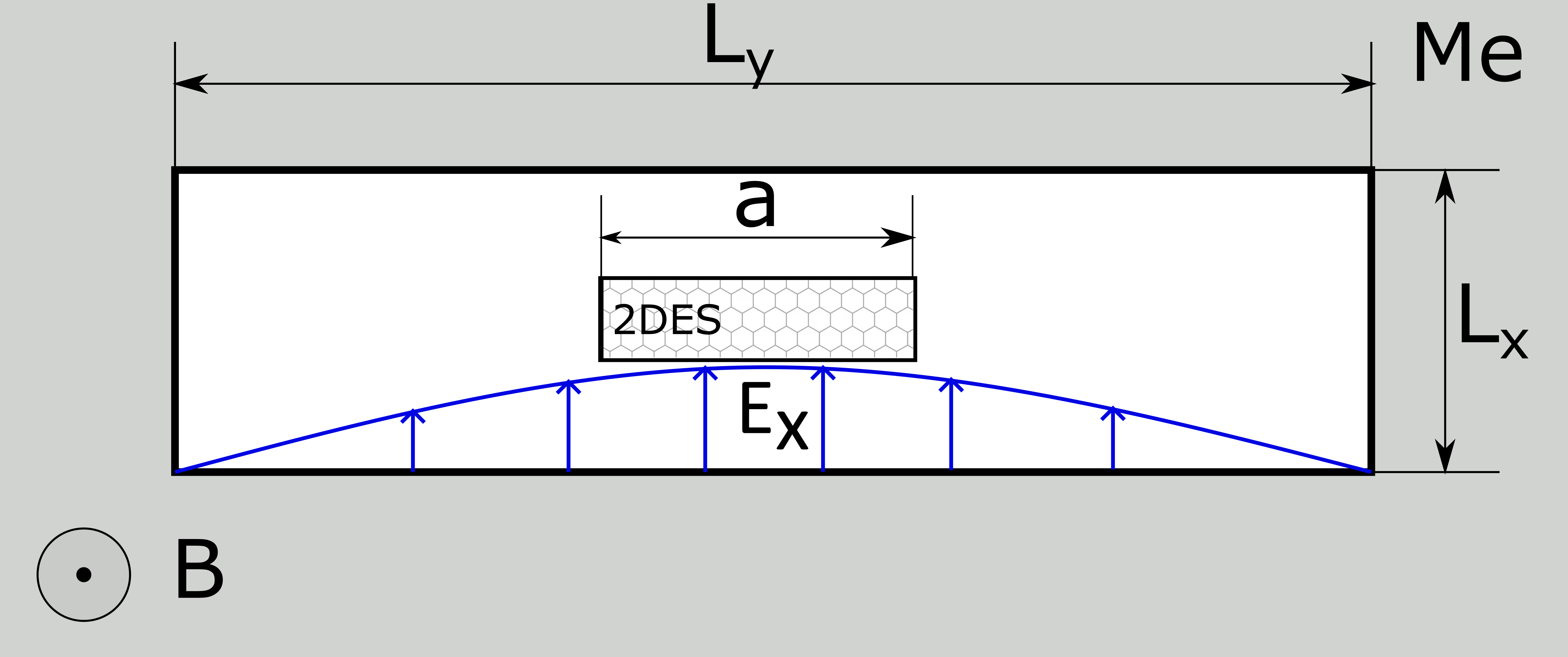}
\caption{Schematic diagram of the system under study. The solution took into account modes polarized along $L_x$ and $L_y$ due to the location of the two-dimensional electrical system at the resonator point, where all modes are polarized along one of these directions. The two-dimensional electron system is small compared to the wavelength of the resonator modes ($a<<\lambda_{res}$), which leads to a uniform field of virtual photons}
\label{fig:Resonator}
\end{figure}

In this article, we solve the problem of the self-energy levels of a two-dimensional electron system in a constant magnetic field in a multimode resonator (Fig. \ref{fig:Resonator}). To do this, we reduce the Hamiltonian of the system to a form similar to a multidimensional harmonic oscillator with a potential of a quadratic form. We then diagonalize this form using the self-energy method for the eigenmodes of a resonator with polarization along one axis. This allows us to reduce the corresponding block of the potential matrix. We represent the reduced matrix as the sum of the diagonal matrix and the direct product of the coupling vector of the magnetic field with photon modes of perpendicular polarization (\ref{eq:mag_link}). This allows us to reduce the equation for finding the eigenfrequencies of the system to the form (\ref{eq:Ditto-finish}). We eliminate the logarithmic divergence that arises when summing the series, restricting ourselves to modes for which the dipole approximation is applicable within the boundaries of a two-dimensional electron system. Here we will show that taking into account a large number of modes leads to a result that is fundamentally different from the single-mode case in that, for a sufficiently high mode density and the strength of photon interaction, it is possible to obtain an arbitrarily small value of the natural frequency for the lowest natural frequency of the system at a fixed cyclotron frequency.

In this paper, we separate the system under study into a system of independent harmonic oscillators and find the eigenfrequencies of the states. This allows us to account for the influence of resonator modes with arbitrarily specified characteristics, adding only the requirement that the dipole approximation be applicable and that the two-dimensional electron system be located at a high symmetry point of the resonator.

We begin the solution by writing out the Hamiltonian of a system consisting of an electron in a two-dimensional system with a magnetic field perpendicular to the system and a resonator surrounding the system.

\begin{equation}
    \hat{H}=\hat{H}_e+\hat{H}_{ph}
\end{equation}

But the vector potential of photons also modifies the generalized momentum of the electron and enters into $\hat{H}_e$ when replacing ${\bf\hat{p}}\rightarrow {\bf\hat{p}}+e{\bf \hat{A}}$

\begin{multline}
\label{eq:og_Hamiltonian_0}
    \hat{H}=\frac{[\hat{p}_x+e(\sum_{n=1}^{n_{n}} \hat{A}_{q,n,x})]^2}{2m_e}+\\+\frac{(\hat{p}_y+e\hat{A}_{cl,y}+e(\sum_{n=1}^{n_{max}} \hat{A}_{q,n,y}))^2}{2m_e}+\\+\hbar \sum_{n=1}^{n_{max}} \omega_{ph,n}(\hat{b}_n^+ \hat{b}_n+\frac{1}{2})
\end{multline}

Here ${\bf\hat{A}_{cl}}=-\boldsymbol{e}_yB\hat{x}$ is the vector potential of the magnetic field, where $B$ is the magnetic field strength, ${\hat{A}_{q,n,xy}}$ is the vector potential of photons supported by the resonator modes with the corresponding polarization at the location of the two-dimensional electron system, $\hat{b}_n$ is the ladder operator for the resonator modes. The multimode resonator is represented as a series of harmonic oscillators with standard expressions for the generalized coordinate and momentum:

\begin{equation}
\label{eq:standard}
    \hat{z}=\sqrt{\frac{\hbar}{2m_{ph}\omega_{ph}}}(\hat{b}+\hat{b}^+), \hat{p}_z=i\sqrt{\frac{\hbar m_{ph}\omega_{ph}}{2}}(\hat{b}^+-\hat{b})
\end{equation}

The transition to the momentum representation for modes polarized by $$e_x$$ $$\frac{{\hat{p}}_z}{m_{ph}\omega_{ph}}\longrightarrow{\hat{z}}, m_{ph}\omega_{ph}{\hat{z}}\longrightarrow-{\hat{p}}_z$$ leads to the following form of expressions (\ref{eq:standard}):

\begin{equation}
\label{eq:p-standard}
    {\hat{z}}=i\sqrt{\frac{\hbar}{2m_{ph}\omega_{ph}}}(\hat{b}^+-\hat{b}), {\hat{p}}_z=-\sqrt{\frac{\hbar m_{ph}\omega_{ph}}{2}}(\hat{b}^++\hat{b})
\end{equation}

For a small size of a two-dimensional system $a<<\lambda$, the dipole approximation is applicable, which leads to the possibility of expressing the vector potential of a photon in the mode with x polarization in the form:

\begin{equation}
\label{eq:dipole}
    \hat{\boldsymbol{A_x}}\approx A_0 \boldsymbol{e}_x(\hat{b}^++\hat{b})=-A_0 \boldsymbol{e}_x \sqrt{\frac{2}{m_{ph}\hbar \omega_{ph}}}\hat{p}_z=-\frac{1}{e}\boldsymbol{e}_x \hat{p}_z
\end{equation}

If we take the mass of a photon to be equal to:

\begin{equation}
\label{eg:mass}
    m_{ph}=\frac{2(e A_0)^2}{\hbar \omega_{ph}}=\frac{2 e^2}{\epsilon_0 V \omega_{ph}^2}
\end{equation}

For photons in y-polarized modes we obtain:

\begin{equation}
    \label {eq_y_dipole}
    \hat{\boldsymbol{A_y}}\approx A_0 \boldsymbol{e}_y(\hat{b}^++\hat{b})=A_0 \boldsymbol{e}_y \sqrt{\frac{2m_{ph}\omega_{ph}}{\hbar}}\hat{z}=B \boldsymbol{e}_y \hat{z}
\end{equation}
This leads us to a Hamiltonian of the form:

\begin{multline}
\label{eq:og_Hamiltonian}
    \hat{H}=\frac{[\hat{p}_x-\sum_{n=1}^{n_{max}} \hat{p}_{z,n,x}]^2+(\hat{p}_y+eB[\hat{x}+\sum_{n=1}^{n_{max}} \hat{z}_{n,y}])^2}{2m_e}+\\
    +\hbar \sum_{n=1}^{n_{max}} \left(\frac{\hat{p}_{z,n,x}^2}{2m_{ph,n,x}}+\frac{m_{ph,n,x} \omega_{n,x}^2 \hat{z}_{n,x}^2 }{2}\right)+\\+\sum_{n=1}^{n_{max}}\left(\frac{\hat{p}_{z,n,y}^2}{2m_{ph,n,y}}+\frac{m_{ph,n,y} \omega_{n,y}^2 \hat{z}_{n,y}^2 }{2}\right)
\end{multline}

After switching to the momentum representation with respect to the y-axis $\hat{p}_y=-eB\hat{y}$, we perform a series of transformations to make the particle mass isotropic with respect to the momentum operators.

To do this, for modes polarized along the x-axis, we perform a transformation of the form:

\begin{equation}
\label{eq:first_transform}
    \hat{\bf{p'}}=\begin{pmatrix}
        1 & -1 &...& & -1\\
        0 & \sqrt{\frac{m_e}{m_{ph,1,y}}} & 0 & ... & 0\\
        \vdots & 0 & \ddots\\
         & \vdots \\
         0 & & \dots & & \sqrt{\frac{m_e}{m_{ph,n,y}}}
    \end{pmatrix} \hat{\bf{p}}
\end{equation}

In this case, for modes with polarization relative to y, the following replacement is made:
\begin{equation}
    \label{eq:y_transform}
    \hat{p}'_{z,n,y}=\sqrt{\frac{m_e}{m_{ph,n,y}}}p_{z,n,y}
\end{equation}

This leads us to a Hamiltonian of the form:
\begin{multline}
\label{eq:Hamiltonian2}
    \hat{H}=\frac{\hat{p}_x^2+\sum_{n=1}^{n_{max}} \hat{p}_{z,n,x}^2+\sum_{n=1}^{n_{max}} \hat{p}_{z,n,y}^2}{2m_e}+\\+\frac{eB}{2m_e}(\hat{x}-\hat{y}+\sum_{n=1}^{n_{max}} \sqrt{\frac{m_{ph,n,y}}{m_e}}\hat{z}_{n,y}])^2+\\+ \sum_{n=1}^{n_{max}} \frac{\omega_{ph,n}^2}{2}(\sqrt{m_e}\hat{z_n}-\sqrt{m_{ph,n}}\hat{x})^2+\sum_{n=1}^{n_{max}} \frac{\omega_{ph,n}^2}{2}(\sqrt{m_e}\hat{z_n})^2
\end{multline}

Finally, we get rid of the y-coordinate in the Hamiltonian and obtain the final form of the Hamiltonian.

\begin{multline}
\label{eq:final}
    \hat{H}=\frac{\hat{p}_x^2+\sum_{n=1}^{n_{max}} \hat{p}_{z,n,x}^2+\sum_{n=1}^{n_{max}} \hat{p}_{z,n,y}^2}{2m_e}+\\+\frac{eB}{2m_e}(\hat{x}+\sum_{n=1}^{n_{max}} \sqrt{\frac{m_{ph,n,y}}{m_e}}\hat{z}_{n,y}])^2+\\+ \sum_{n=1}^{n_{max}} \frac{\omega_{ph,n}^2}{2}(\sqrt{m_e}\hat{z_n}-\sqrt{m_{ph,n}}\hat{x})^2+\sum_{n=1}^{n_{max}} \frac{\omega_{ph,n}^2}{2}(\sqrt{m_e}\hat{z_n})^2
\end{multline}

This expression is the Hamiltonian of an n+1-dimensional harmonic oscillator, where the potential is given by a quadratic form of the form:
\begin{equation}
\hat{\Lambda}_p=\begin{pmatrix}
        \hat{x} &
        \hat{z}_{n,y} &
        \hat{z}_{n,x}
    \end{pmatrix}\hat{\Lambda}\begin{pmatrix}
        \hat{x} \\
        \hat{z}_{n,y} \\
        \hat{z}_{n,x}
    \end{pmatrix}:
\end{equation}

\begin{equation}
\label{eq:potential}
\hat{\Lambda}=
\begin{pmatrix}
        \frac{(eB)^2}{2m_e}+\sum_{n=1}^{n_{max}}\frac{m_{ph,n}\omega_{ph,n}^2}{2} & \hat{\tau}_y & \hat{\tau}_x \\
        \hat{\tau}^+_y & \hat{\Lambda}_{ph,y} & 0 \\
        \hat{\tau}^+_x & 0 & \hat{\Lambda}_{ph,x}
    \end{pmatrix}   
\end{equation}

Where the blocks $\Lambda_{ph,xy}$ and $\tau_{xy}$ are defined using the expressions:
\begin{equation}
\label{eq:Eph}
    \hat{\Lambda}_{ph,x}=\begin{pmatrix}
        \frac{m_e \omega^2_{ph,1,x}}{2} & & 0\\
         & \ddots & \\
         0 & & \frac{m_e \omega^2_{ph,N,x}}{2} \\
    \end{pmatrix}
\end{equation}
\begin{equation}
\label{eq:Eph_y}
    \hat{\Lambda}_{ph,y}=\begin{pmatrix}
        \frac{m_e \omega^2_{ph,1,y}}{2}+\frac{B^2 e^2}{2 m_{ph,1,y}} & & \frac{B^2 e^2}{2 \sqrt{m_{ph,1,y}m_{ph,N,y}}}\\
         & \ddots & \\
         \frac{B^2 e^2}{2 \sqrt{m_{ph,1,y}m_{ph,N,y}}} & & \frac{m_e \omega^2_{ph,N,x}}{2}+\frac{B^2 e^2}{2 m_{ph,N,y}} \\
    \end{pmatrix}
\end{equation}
\begin{equation}
\label{eq:tau}
    \hat{\tau}_x=\left(\frac{\omega_{ph,1,x}e\sqrt{ m_e  }}{\sqrt{2\varepsilon_0 V}},...,\frac{\omega_{ph,n}e\sqrt{ m_e  }}{\sqrt{2\varepsilon_0 V}}\right)
\end{equation}

\begin{equation}
\label{eq:tau_y}
    \hat{\tau}_y=\left(\frac{B^2 e^2}{2 \sqrt{m_{e}m_{ph,1,y}}},\dots,\frac{B^2 e^2}{2 \sqrt{m_{e}m_{ph,N,y}}}\right)
\end{equation}

This allows us to represent the problem as finding the eigenfrequencies of n+1 independent harmonic oscillators with mass $m_e$. Thus, we can find the eigenvalues of the Hamiltonian (\ref{eq:og_Hamiltonian_0}), which depend on the occupation numbers as follows:

\begin{equation}
\label{eq:energy}
    E(n_1,...,n_N)=\sum_{i=1}^{N}\hbar\omega_{i}n_i
\end{equation}

The natural frequencies are related to the natural values of the potential matrix (\ref{eq:potential}) by an expression of the form: 

\begin{equation}
    \label{eq:lambda}
    \omega_i=\sqrt{\frac{2\lambda_i}{m_e}}
\end{equation}

Using the self-energy method discussed in ~\cite{datta2005quantum}, we can reduce the matrix by adding the blocks $\hat{\Lambda}_{ph,x}$ and $\hat{\tau}_x$ as an addition to the first diagonal element of the matrix. Unlike the usual application of the self-energy method, we apply this approach not to the Hamiltonian of the system, but to the potential matrix, which has a form similar to the Hamiltonian of systems for which the self-energy approach is applied. After substituting the expression for the photon mass (\ref{eg:mass}), the first diagonal element of the reduced matrix takes the form:

\begin{equation}
    \label{eq:Datta_0}
    \Lambda_{r}=\frac{(eB)^2}{2m_e}+\frac{n_{max}e^2}{\varepsilon_0 V}  - \Sigma
\end{equation}

Where $\Sigma = \tau^+ (\Lambda - \lambda E)^{-1}\tau$, and E is the identity matrix.

Furthermore, we note that the reduced matrix can be represented as the sum of a diagonal matrix and the direct product of the coupling vector of the magnetic field to the resonator mode field with itself:

\begin{multline}
    \label{eq:rank_up}
    \Lambda=Diag(\Lambda_r-\frac{(eB)^2}{2m_e},\Lambda_{ph,y,1}-\frac{(eB)^2}{2m_{ph,y,1}},\\,\dots,\Lambda_{ph,y,N}-\frac{(eB)^2}{2m_{ph,y,N }})+uu^+
\end{multline}
, where the vector u has the form:

\begin{equation}
\label{eq:mag_link}
u=\begin{pmatrix}
\sqrt{\frac{m_e \omega_c^2}{2}}\\
\sqrt{\frac{B^2 e^2}{2 m_{ph,1,y}}} \\
\vdots\\
\sqrt{\frac{B^2 e^2}{2 m_{ph,N,y}}}
\end{pmatrix} 
= 
\begin{pmatrix}
    \sqrt{\frac{m_e \omega_c^2}{2}}\\
    \sqrt{\frac{B^2 \varepsilon V}{2}}\\
    \vdots\\
    \sqrt{\frac{B^2 \varepsilon V}{2}}
\end{pmatrix}
\end{equation}
This allows us to use the method of finding the eigenvalues of a matrix after updating the first rank, known in linear algebra \cite{golub1973some}. In this case, the equation for the eigenvalues of the system takes the form:

\begin{equation}
\label{eq:Ditto}
    1+\frac{m_e\omega_c^2}{2(\Lambda_r-\lambda)}+\sum_{n=1}^{n_{max}}\frac{u_n^2}{\Lambda_{ph,n,y}-\lambda}=0
\end{equation}

Using the expressions for $\lambda$ (\ref{eq:lambda}), $\Lambda_{xy}$ (\ref{eq:Eph},\ref{eq:Eph_y}), and $\tau$ (\ref{eq:tau},\ref{eq:tau_y}) and dividing both sides of the equation by $\frac{m_e}{2}$, we arrive at the expression

\begin{multline}
\label{eq:Ditto-translate}
    \omega_c^2- \omega_i^2 -\frac{2e^2}{m_e\varepsilon_0 V} \sum_{n=1}^{n_{max}}[1-\frac{\omega_{ph,n,x}^2}{\omega_{ph,n,x}^2-\omega_i^2}]-\\- \frac{2e^2}{m_e\varepsilon_0 V}\sum_{n=1}^{n_{max}}\frac{\omega_{ph,y,n}^2}{\omega_{ph,n,y}^2-\omega_i^2}-\\-\frac{4e^4}{m_e^2\varepsilon_0^2 V^2}\sum_{n=1}^{n_{max}}\frac{\omega_{ph,y,n}^2}{\omega_{ph,n,y}^2-\omega_i^2}\sum_{n=1}^{n_{max}}[1-\frac{\omega_{ph,n,x}^2}{\omega_{ph,n,x}^2-\omega_i^2}]=0
\end{multline}

Next, we introduce the coupling constant $\chi^2_{xy}=\frac{2e^2}{m_e \varepsilon_0 V \omega_{ph,1,xy}^2}$ and the resonator mode numbering function $g_{xy}(n)=\frac{\omega_{ph,n,xy}}{\omega_{ph,1,xy}}$, which shows the relationship of the resonator mode frequency with the corresponding number relative to the lowest resonator mode. We also replace sums of the form $S_{xy}=\sum_{n=1}^{n_{max}}\frac{1}{g_{xy}^2(n)-\frac{\omega^2_i}{\omega_{ph,1,xy}^2}}$. Using these, we obtain the equation:

\begin{multline}
\label{eq:Ditto-finish}
    \omega_c^2- \omega_i^2(1 +\chi_x^2 S_x+\chi_y^2 S_y+\chi_x^2\chi_y^2S_x S_y)=0
\end{multline}

This equation shows that for small values of the coupling constant, the natural frequencies of the system approach the cyclotron frequency and the natural frequencies of the resonator due to the zeroing of all terms in the parentheses except unity.

Also, the form of the terms in the summation shows the condition for the convergence of the series $g(n)\sim n^{\frac{1}{2}+\delta},\delta>0$. This implies a logarithmic divergence of the series when summed over two indices, which corresponds to the case of a rectangular resonator. To eliminate this divergence, we restrict ourselves to considering only modes with indices no higher than $N_{dip,xy}=\frac{L_{xy}}{a}$, since for higher modes the requirement of field uniformity, without which the dipole approximation ceases to work, is no longer satisfied.

The case of a resonator with one side extended relative to the other deserves special mention, so that $min(N_{dip,x},N_{dip,y})=1$. Thus, the numbering function has the form of a linear function, which allows summation to infinity over one of the indices, since higher values of the second index no longer satisfy the conditions for the applicability of the dipole approximation:

\begin{equation}
\label{eq:equidistant}
    g(n)= 1+d(n-1)
\end{equation}
In this case, the equation (\ref{eq:Ditto-finish}) takes the form

\begin{equation}
\label{eq:Ditto-finish-2}
    \omega_c^2- \omega_i^2 -\chi^2\omega_i^2 \sum_{n=1}^{n_{max}}[\frac{1}{(1+d(n-1))^2-\frac{\omega_i^2}{\omega_{ph,1}^2}}]=0
\end{equation}

In this case, when $\frac{\omega^2_i}{\omega_{ph,1}^2}<<1$, equation (\ref{eq:Ditto}) can be expanded in powers of $\omega^2_i$. When expanding to the first power of smallness, we obtain:

\begin{multline}
\label{eq:Taylor}
    \omega_c^2- \omega_i^2 -\chi^2\sum_{n=1}^{n_{max}}\frac{\omega^2_i}{2(1+d(n-1))^2}]=0
\end{multline}

The solution of which yields the value of the lowest eigenvalue of the potential. Since the condition $\omega_{-}^2<min(\omega_c^2,\chi^2\omega_{ph,1}^2)$ is always satisfied for the lowest potential value,

Thus, when the above condition is satisfied, the value of the lowest eigenfrequency of the system under study is:

\begin{equation}
\label{eq:omega}
     \omega_{min} = \omega_c\sqrt{\frac{1 }{1+\frac{4\chi^2}{d^2}\sum_{n=d}^{n_{max}}{\frac{1}{n^2}}}}
\end{equation}

And, with a large number of photonic modes:

\begin{equation}
\label{eq:limit}
    lim_{n\xrightarrow{}\infty} \omega_{min} = \omega_c\sqrt{\frac{1 }{1+\frac{4\chi^2}{d^2}(\frac{\pi^2}{6}-\sum_{n=1}^{d-1} \frac{1}{n^2})}}
\end{equation}

\begin{figure}[ht!]
\includegraphics[width=0.9\linewidth]{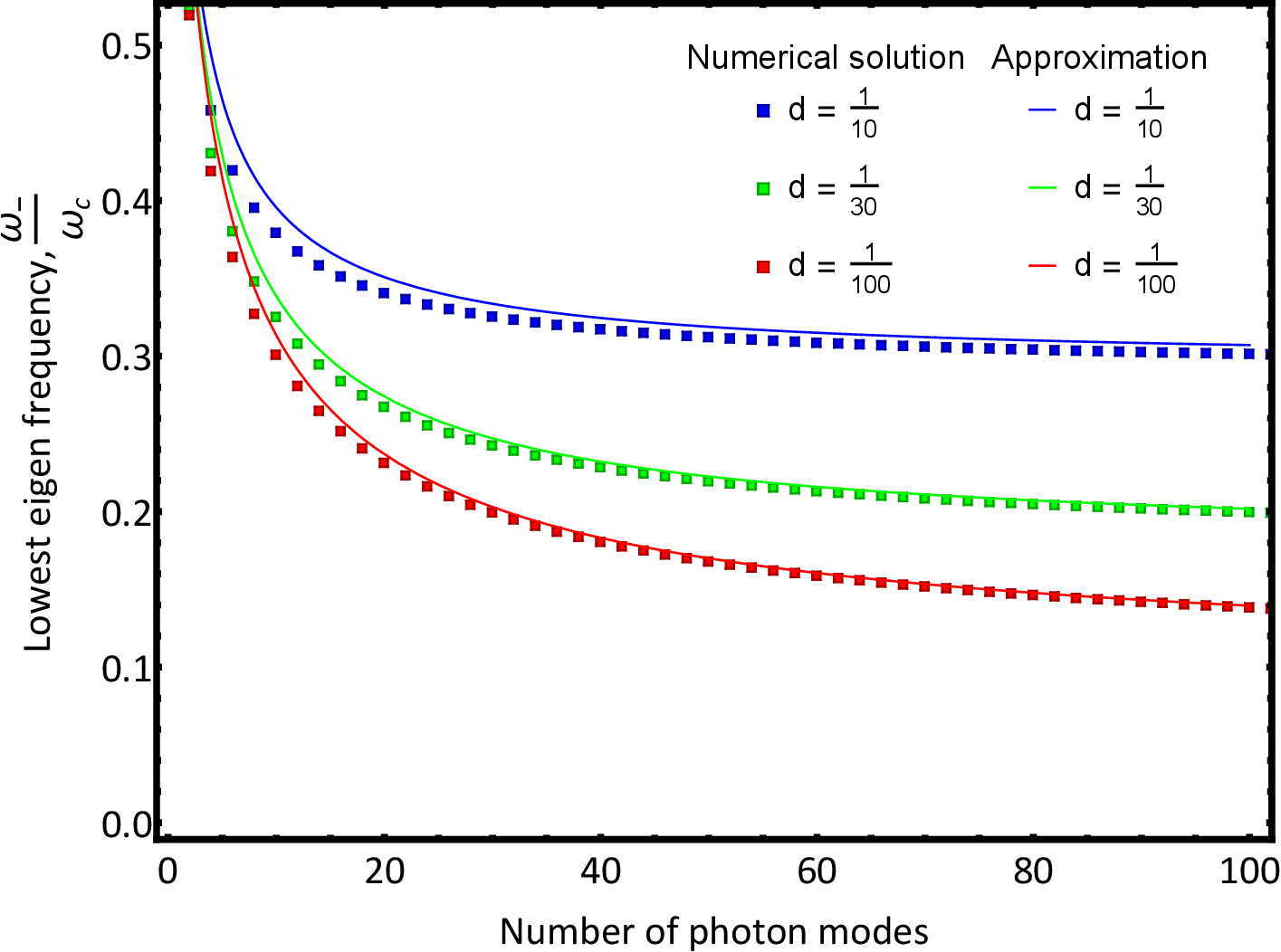}
\caption{Plots of level positions versus the number of photons for various distances between the resonator mode eigenfrequencies. Numerical solution (dots) and approximate analytical solution (\ref{eq:omega}) (solid lines)}
\label{fig:Mode_num}
\end{figure}

The graph (\ref{fig:Mode_num}) shows the relationship between the lowest eigenfrequency positions for various mode densities as a function of the resonator mode number, as found by numerically solving equation (\ref{eq:Ditto}) and by solving the approximate equation (\ref{eq:Taylor}).

It is evident that the accuracy of the approximate equation's solution increases with increasing mode density and the number of photons. The obtained dependences indicate that increasing the number of modes without increasing their density allows a limited decrease in the position of the system's lowest eigenlevel. However, increasing the mode density or increasing the photon mass allows the system's eigenfrequency to be positioned arbitrarily low.

\begin{figure}[ht!]
\includegraphics[width=0.9\linewidth]{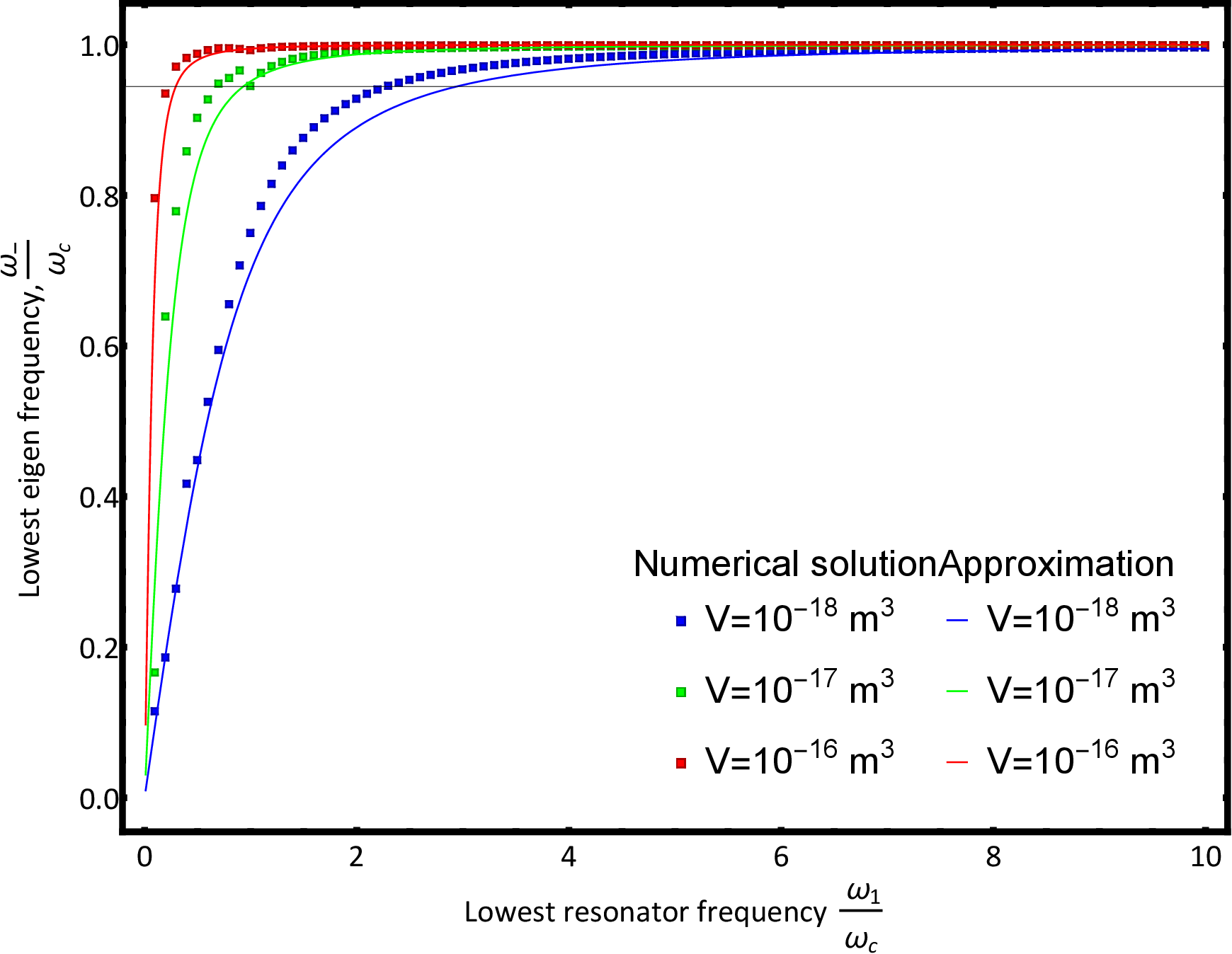}
\caption{Position of the lower polariton level as a function of the frequency of the lowest photon mode, calculated numerically (dots) and using the approximate analytical formula (\ref{eq:omega}) (solid lines) for various resonator volumes.}
\label{fig:limit}
\end{figure}

We also calculated how the position of the lower mode relative to the Landau level changes for several sets of physical values for a resonator with an extremely large number of modes (Fig. \ref{fig:limit}).

In the limit of strong photon interactions, the approximate formula (\ref{eq:limit}) demonstrates a linear dependence of the system's natural frequency on the magnitude of the zero-point oscillation vector potential. This dependence demonstrates the inapplicability of perturbation theory in this limit, since the contribution of the lowest degree of smallness of the vector potential obtained with this approach is quadratic.

Another detail of the limit approximation is the nontrivial dependence of the coupling strength on the resonator shape. Increasing the resonator volume leads to a decrease in the coupling strength, but decreasing the resonator thickness leads to an increase in the intermode spacing. Thus, the effect studied has the greatest impact on two-dimensional systems with a sufficient thickness to ensure a high mode density and a small area that will not weaken the photon interaction.

In this paper, we proposed a model for accounting for the influence of a multimode resonator on the Landau levels in a two-dimensional system and demonstrated the possibility of separating the system into a set of independent harmonic oscillators. We also discovered the possibility of softening the polariton to zero energy with increasing photon coupling strength and the resonator mode density. We used the dipole approximation of the vector potential for the resonator mode fields and the assumption that the two-dimensional electron system is small compared to the resonator mode wavelength. To diagonalize the potential matrix, the self-energy method and the method of finding the eigenvalues of the matrix after updating the first rank were applied.

\subsection{Acknowledgments}

This work was funded by the Basis Foundation for the Development of Theoretical Physics and Mathematics. No additional grants were received for conducting or supervising this specific research.

\subsection{Conflict of Interest}
The authors of this work declare that they have no conflicts of interest.

\bibliography{sample}

\end{document}